\documentclass{article}
\usepackage{graphicx,color}
\usepackage{cite}
\usepackage{amssymb}
\usepackage{amsmath}
\usepackage{epstopdf}
\usepackage[american]{babel}
\usepackage[T1]{fontenc}
\usepackage[utf8]{inputenc}
\usepackage{authblk}
\begin{document}
\begin{titlepage}

\title{On Black Hole Structures in Scalar-Tensor Theories of Gravity}

\author[1,3,4]{Kirill A. Bronnikov\thanks{kb20@yandex.ru}}

\author[2,3]{J\'ulio C. Fabris\thanks{fabris@pq.cnpq.br}}

\author[2]{Denis C. Rodrigues\thanks{deniscr@gmail.com}}

\affil[1]{VNIIMS, Ozyornaya ul. 46, Moscow 119361,
Russia} 

\affil[2]{Departamento de F\'{\i}sica - Universidade
Federal do Esp\'{\i}rito Santo - Brazil}
\affil[3]{National Research Nuclear University
“MEPhI,” Kashirskoe sh. 31, Moscow 115409, Russia}
\affil[4]{ Institute of Gravitation and
Cosmology, PFUR, ul. Miklukho-Maklaya 6, Moscow 117198, Russia}
\maketitle

\begin{abstract}
We review some properties of black hole structures appearing in gravity with a
massless scalar field, with both minimal and nonminimal coupling. The main properties of the resulting cold black holes are described. The study of 
black holes in scalar-gravity systems is extended to $k$-essence theories, and some examples are explicitly worked out. In these cases, even while the existence of horizons is possible, the metric regularity requirement on 
the horizon implies either a cold black type structure or a singular behavior 
of the scalar field.
\end{abstract}

\end{titlepage}

\section{Black holes in general relativity}

The existence of black holes is one of the most important predictions of the General Relativity
theory. Even if some aspects of black hole physics can be obtained from the Newtonian theory,
the complete concept of black holes includes the notion of an event horizon and a causal
structure typical of the relativistic theory. We are perhaps close to a direct detection of
black holes, and the amount of evidence on black holes is accumulating. The recent detection
of gravitational waves by the LIGO experiment\cite{gw} is considered as a clear evidence for the existence of these objects since the measured gravitational wave carries the signature of the coalescence of two very massive black holes.
Moreover, the centers of galaxies may be inhabited by black holes with masses of the order of
millions to billions of solar masses. There are also many indications of the existence
of black holes resulting from stellar evolution, with masses of the order of a few
solar masses.

There are four most well-known black hole solutions in General Relativity: the uncharged
static black hole, given by the Schwarzschild solution; the charged static black hole,
the Reissner-Nordstr\"om solution; the uncharged rotating black hole, the Kerr solution;
and the rotating, charged black hole, the Kerr-Newman solution. The properties of these
solutions led to the so-called {\it no-hair conjecture}: in general, black holes must be characterised by only three parameters in Einstein-Maxwell theory, the mass $M$, the charge $Q$ and the angular
momentum $J$; however, other parameters corresponding to new charges are added if more gauge fields are included into consideration, as happens, e.g., with string theory inspired Lagrangians; also new parameters like magnetic monopole charges may appear when general new gauge fields are taken into account. There is strong evidence that this conjecture may be violated
in some situations, in particular, when new sources like scalar fields are considered
--- see the contributions of Herdeiro et al. in the present volume.

The prototype of the black hole structure is given by the Schwarzschild solution, for
which the line element reads
\begin{eqnarray}
ds^2 = \biggr(1 - \frac{2GM}{r}\biggl)dt^2 - \biggr(1 - \frac{2GM}{r}\biggl)^{-1}dr^2
 - r^2d\Omega^2.
\end{eqnarray}
Here, the crucial concept of an {\it event horizon} appears, through the coordinate
singularity at $r_H = 2GM$.
The Reissner-Nordstr\"om solution is given by a line element with two parameters,
mass and charge, which reads
\begin{eqnarray}
ds^2 =  \biggr(1 - \frac{2GM}{r} + \frac{Q^2}{r^2}\biggl)dt^2
- \biggr(1 - \frac{2GM}{r} + \frac{Q^2}{r^2}\biggl)^{-1}dr^2 - r^2d\Omega^2.
\end{eqnarray}
It brings to scene the notion of a Cauchy horizon, here given by
$r_- = M - \sqrt{M^2 - Q^2}$, while the event horizon appears at
$r_+ = M + \sqrt{M^2 - Q^2}$. The Reissner-Nordstr\"om solution also implies the notion
of an extremal black hole, represented by the configuration where the Cauchy horizon
coincides with the event horizon, $r_+ = r_-$, so that $M = Q$. Moreover, if $Q > M$,
the horizons disappear, and the central singularity becomes accessible to an external
observer: this is a naked singularity. These new features appear also, in a different form,
in the Kerr and Kerr-Newman solutions.

One of the important results in the study of black hole physics is the discovery of
the thermodynamics behavior of those objects. The temperature and entropy of the
Schwarzschild black hole are given by
\begin{eqnarray}
    T_H = \frac{1}{8\pi}\frac{1}{M}, \qquad S_H = 4\pi M^2 = \frac{A_H}{4},
\end{eqnarray}
where $A_H$ is the area of the event horizon surface.
The temperature of the Reissner-Nordstr\"om black hole is given by
\begin{eqnarray}
    T_{RN} &=& \frac{1}{8\pi M}\biggl(1 - 16\pi^2\frac{Q^4}{A^2}\biggl),\\
    A &=& 4\pi R_+^2.
\end{eqnarray}
The temperature of the Reissner-Nordstr\"om black hole becomes zero when the extremal
condition is satisfied, $M = Q$. There is no well-defined temperature for naked singularities.
The entropy of the Reissner-Nordstr\"om black holes is still given by the the area law:
\begin{eqnarray}
    S = \pi r_+^2.
\end{eqnarray}
The entropy is always finite, including the extremal case (see the contribution by Lemos
in the present volume). But, it becomes ill-defined in the naked singularity case. For a
general view of these standard results, see Refs. \cite{Chandra,frolov}

This quick review of the standard results in standard black hole physics settles
a basic framework to the main subject of the present text: the fate of static black hole
solutions when the scalar field are added to the gravitational Lagrangian, with minimal
or non-minimal coupling. Seminal studies in this sense have been presented in
\cite{fisher,ellis,bronnikov}. In some situations, new features appear: in general,
black hole solutions in the presence of a scalar field require a violation of the
standard energy conditions, so that the scalar field must have a phantom nature;
the event horizon surface has an infinite area; the Hawking temperature is always zero.
Essentially, these results have been obtained in Refs. \cite{nelson,kirill}. We review
these results and present some new ones, with a similar structure, obtained in the
context of k-essence models \cite{denis}.

\section{Scalar fields}

The introduction of scalar fields is the simplest extension of General Relativity.
It adds a spin zero degree of freedom to the theory.
We can think on some different possibility to introduce a scalar field in a gravitational
context. The first one makes use of the minimal coupling between the scalar field
and gravitation, and it is defined by the Lagrangian
\begin{eqnarray}
{\cal L} = \sqrt{-g}\biggr\{R - \epsilon\phi_{;\rho}\phi^{;\rho} + 2V(\phi)\biggl\}.
\end{eqnarray}
In this Lagrangian, $V(\phi)$ is the self-interaction potential. This kind of theory in
general satisfies the energy conditions if $\epsilon = 1$ (an ordinary field) and violates the
energy conditions if $\epsilon = - 1$ (a phantom fields).

We may have a much less trivial configuration using a non-minimal coupling,
\begin{eqnarray}
{\cal L} = \sqrt{-g}\biggr\{f(\phi)R - \omega(\phi)\frac{\phi_{;\rho}\phi^{;\rho}}{\phi} + 2V(\phi)\biggl\}.
\end{eqnarray}
where $f(\phi)$ and $\omega(\phi)$ are, in general, functions of the scalar field.
The case
\begin{eqnarray}
    f(\phi) = \phi, \qquad \omega(\phi) =\mbox{constant}, \qquad V(\phi) = 0,
\end{eqnarray}
defines the traditional Brans-Dicke theory, the prototype of the scalar-tensor theories.

Another possibility of including a scalar field is a non-canonical kinetic term,
leading to $k$-essence models,
\begin{eqnarray}
    {\cal L} = \sqrt{-g}\biggr\{R - \omega(\phi)f(X)+ 2V(\phi)\biggl\},
\end{eqnarray}
where
\begin{eqnarray}
    X = \phi_{;\rho}\phi^{;\rho},
\end{eqnarray}
and $f$ is a given function of $X$.

Other possibilities are the Horndeski theory \cite{horndesky}, with the most general
scalar-tensor Lagrangian leading to second-order differential equations, and
the Galileon theory \cite{galleon}, where a scalar field exhibits some special translational
symmetry. One example of such a generalization is the theory defined by
\begin{eqnarray}
    {\cal L} = \sqrt{-g}\biggr\{R - G^{\mu\nu}\phi_{;\mu}\phi_{;\nu}+ 2V(\phi)\biggl\},
\end{eqnarray}
where $G_{\mu\nu}$ is the Einstein tensor.

\section{Black holes in minimally coupled models}

In the ordinary case where the scalar field is minimally coupled to gravity, without
a potential, represented by the equations
\begin{eqnarray}
    R_{\mu\nu} - \frac{1}{2}g_{\mu\nu}R &=& \phi_{;\mu}\phi_{;\nu}
    - \frac{1}{2}g_{\mu\nu}\phi_{;\rho}\phi^{;\rho}, \\
    \Box\phi &=& 0,
\end{eqnarray}
there is no black hole solution.
Indeed, let us write the four-dimensional spherically symmetric, static metric as
\begin{eqnarray}
    ds^2 = e^{2\gamma}dt^2 - e^{2\alpha}du^2 - e^{2\beta}d\Omega^2,
 \end{eqnarray}
where $\alpha$, $\beta$ and $\gamma$ are functions only of the radial coordinate $u$.
The field equations are
\begin{eqnarray}
- 2\beta'' - 3\beta'^2 + 2\beta'\alpha'+ e^{2(\alpha - \beta)}
        &=& \frac{\epsilon}{2}\phi'^{2},  \\
2\gamma'\beta' + \beta'^2 - e^{2(\alpha - \beta)}
        &=& \frac{\epsilon}{2}\phi'^{2},  \\
\beta''+ \gamma''+ \beta'(\gamma' + \beta'- \alpha') + \gamma'^2 - \alpha'\gamma'
        &=& - \frac{\epsilon}{2}\phi^{2},  \\
    \phi'' + [\gamma' + 2\beta' - \alpha']\phi'&=& 0.
\end{eqnarray}
For an ordinary scalar field, $\epsilon = 1$. Choosing the coordinate condition
\begin{eqnarray}
    \alpha = \gamma + 2\beta,
\end{eqnarray}
  the scalar field equation can be easily solved:
\begin{eqnarray}
    \phi = \phi_0 u + \phi_1.
\end{eqnarray}
    The remaining equations simplify to
\begin{eqnarray}
    - 2\beta'' + {\beta'}^2 + 2\beta'\gamma'+ e^{2(\gamma + \beta)}
            &=& \frac{\epsilon}{2}\phi'^{2},  \\
    2\gamma'\beta' + \beta'^2 - e^{2(\gamma + \beta)}
            &=& \frac{\epsilon}{2}\phi'^{2},  \\
    \beta''+ \gamma''+ {\beta'}^2 - 2\beta'\gamma'
            &=& - \frac{\epsilon}{2}\phi^{2}.
\end{eqnarray}
A combination of those equations leads to
\begin{eqnarray}
    \gamma'' = 0.
\end{eqnarray}
The solution is:
\begin{eqnarray}
    \gamma = bu.
\end{eqnarray}
Identifying,
\begin{eqnarray}
e^{- 2ku} = 1 - 2\frac{k}{x},
\end{eqnarray}
the final metric is
\begin{eqnarray}
\label{f1}
    ds^2 = P^adt^2 - P^{-a}dx^2 - P^{1 - a}x^2d\Omega^2,
\end{eqnarray}
with
\begin{eqnarray}
\label{f2}
    P(x) = 1 - 2\frac{k}{x}, \quad && \phi = - \frac{C}{2k}\ln P(x),  \\
  \label{k}  \frac{m}{k} = a ,\quad && \quad a^2 = 1 - \frac{C^2}{2k^2}.
\end{eqnarray}
In the expressions above, $m$ is the ADM mass, $C$ is a scalar charge in the sense that $\phi \approx C/x$ at large $x$ (but with no universal conservation law in the present case) and 
$k$ is their combination according to (\ref{k}). The metric (\ref{f1}), together with the definitions (\ref{f2},\ref{k}), corresponds to Fisher's solution \cite{fisher}.
Clearly, there is a naked singularity at $x = 2k$: it is a center.

The situation changes drastically if the null energy condition is violated \cite{nelson}.
The null energy condition requires that
\begin{eqnarray}
    \rho + p \geq 0.
\end{eqnarray}
In order to violate this condition, the kinetic term must appear with the {\it wrong} sign.
 If $\epsilon = - 1$ (a phantom field), we have
\begin{eqnarray}
    ds^2 &=& P^adt^2 - P^{-a}dx^2 - P^{1 - a}x^2d\Omega^2, \\
    a^2 &=& 1 + \frac{C^2}{2k^2} > 1.
\end{eqnarray}
This case represents a black hole with an event horizon at $x = 2k$ with two remarkable
features: an infinite event horizon area and zero surface gravity.

\section{Nonminimal coupling}

We can write a general action with a non-minimal coupling without a potential as
\begin{eqnarray}
    {\cal L} = \sqrt{- g}\biggr\{f(\phi) R - \omega(\phi)\frac{\phi_{;\rho}\phi^{;\rho}}
    {\phi}\biggl\}.
\end{eqnarray}
It is not, however,  the most general (the Horndeski theory is the most general one
leading to second-order equations), but it covers a large sample of theories.
Let us fix
\begin{eqnarray}
    f(\phi) = \phi.
\end{eqnarray}
Performing a conformal transformation
\begin{eqnarray}
    g_{\mu\nu} = \phi^{-1}\tilde g_{\mu\nu},
\end{eqnarray}
and writing
\begin{eqnarray}
    \frac{d\phi}{d\sigma} = \biggr|\frac{3 + 2\omega}{\phi^2}\biggl|^{1/2},
\end{eqnarray}
we end up with
\begin{eqnarray}
{\cal L} = \sqrt{- g}\biggr\{R - \epsilon\sigma_{;\rho}\sigma^{;\rho}\biggl\}, \qquad
    \epsilon = \pm 1.
\end{eqnarray}

A class of static, spherically symmetric solutions is given by
\begin{eqnarray}
    ds^2 = P^{-\xi}\biggr\{P^adt^2 - P^{-a}dx^2 - P^{1 - a}x^2d\Omega^2\biggl\},
\end{eqnarray}
with
\begin{eqnarray}
    P(x) = 1 - 2k/x,&& \quad \phi = P^\xi,  \\
    b/k = a, \quad &&  \quad a^2 = 1 - (3 + 2\omega)\xi^2, \\
    2k^2\mbox{sign} k &=& 2b^2 + \epsilon C^2.
\end{eqnarray}
Black holes are only possible when $3 + 2\omega < 0$, corresponding to the phantom
configuration in the Einstein's frame.

Are the solutions described above, with a minimal or non-minimal coupling, black holes?
They have some striking features:
\begin{enumerate}
\item
Their surface gravity is zero. It means that their Hawking temperature (if it is possible
to define it!) is zero. For this reason, these black holes has been called
{\bf cold black holes}.
\item
The area of the event horizon is infinite. This means an infinite entropy if we follow
the area law. But if the temperature is zero, the entropy cannot be in principle infinite.
Hence, the computation of the entropy in this case may not follow the classical lines
fixed for black holes emerging in General Relativity.
\item
The tidal forces are infinite at the event horizon hypersurface. But point particles
can safely cross this hypersurface.
\item
The solutions are asymptotically flat. Hence, static observers at infinity can be defined.
\end{enumerate}
There is another important feature: The solutions contain a scalar charge, hence these
solutions can be called hairy black holes.
This happens even in the Einstein frame if the energy conditions are violated.
This fact points at a limitations of the so-called {\it no-hair theorem}, which seems
to be restricted to very specific situations.

In fact, the general solution for these scalar black holes reveals two types of
asymptotically flat black hole, besides the trivial (Schwarzschild) one \cite{kirill}:
Type B1 black holes, where the horizon can be crossed in a finite proper time by an
infalling particle; Type B2 black hole, where the horizon is reached in a infinite
proper time for an infalling particle.
In both cases, there is always a throat on the way from spatial infinity to the horizon.

\section{Stability}

The usual GR black hole solutions are stable \cite{Chandra}. These Cold Black Holes, are they stable? 
The stability of black holes is a very delicate subject, technically and conceptually. Technically because 
it is almost impossible to integrate the perturbed equations even in the simplest case, that
of pure radial perturbations; conceptually because the results may be different for different perturbation 
formalisms.

Let us consider just radial perturbations:
\begin{eqnarray}
	\alpha = \alpha_0(u) + \delta\alpha(u,t),&& \quad \beta = \beta_0(u) + \delta\beta(u,t),  \\
	\quad \gamma = \gamma_0(u) &+& \delta\gamma(u,t),
\end{eqnarray}
where $\alpha_0$, $\beta_0$ and $\gamma_0$ are the previously found background solutions and
$\delta\alpha$, $\delta\beta$ and $\delta\gamma$ are (time-dependent!) perturbations around them.
In analyzing perturbations around the Schwarzschild solution, for example, it is very convenient to 
choose the coordinate condition $\delta \beta = 0$.
However, here the situation is more delicate since there is a throat in the solutions outside the horizon, 
the condition above turns out to be a {\it physical condition}: only perturbations that do not affect the 
throat are considered.

It is possible, for Cold Black Holes, to choose a coordinate condition similar to that of the background.
\begin{eqnarray}
	\delta\alpha = \delta\gamma + 2\delta\beta.
\end{eqnarray}
The perturbed equations then may be reduced to a single master equation connected with the perturbation 
of the scalar field:
\begin{eqnarray}
               e^{4\beta(u)}\delta\ddot\phi - \delta\phi''= 0.
\end{eqnarray}
The primes mean derivative with respect to $u$.
This equation can be reduced to
\begin{eqnarray}
	\delta\phi'' + e^{4\beta(u)}\omega^2\delta\phi = 0.
\end{eqnarray}

The condition of stability is fixed as follows: specify the boundary conditions at the horizon and at infinity;
verify if there are solutions connecting these boundary conditions with $\omega^2 < 0$; if there are solutions satisfying the above conditions at both ends, we conclude that there are divergent growing modes, and
the background static solution is unstable, if not, the background solution is stable.
Following this procedure, we have found the following results \cite{kirill}: the cold black hole solutions
are stable; solutions with a naked singularity are unstable.

Let us consider now a scalar field and metric perturbations that transform under the radial reparametrisation $\Delta u$ as
\begin{eqnarray}
	\delta\phi &\rightarrow& \bar{\delta\phi} = \delta\phi + \phi'\Delta u,  \\
	\delta r &\rightarrow& \bar{\delta r} = \delta r + r'\Delta u,
\end{eqnarray}
where we have defined $r(u) = e^{\beta(u)}.$
The combination
\begin{eqnarray}
	\Psi = r'\delta\phi - \phi'\delta r,
\end{eqnarray}
is invariant under that coordinate transformation. The final equation is \cite{zhidenko}
\begin{eqnarray}
                   e^{2(\alpha - \gamma)} \delta\ddot\phi -  \delta\phi'' - \delta\phi'(\gamma'
			+ 2\beta'- \alpha') + U\delta\phi = 0, \\
                U  \equiv e^{2\alpha}\biggr\{\epsilon(V - e^{-2\beta})\frac{{\phi'}^2}{{\beta'}^2} 
			+ 2 \frac{\phi'}{\beta'}V_\phi + \epsilon V_{\phi\phi}\biggl\},
\end{eqnarray}
where $V$ is a possible potential according to the Lagrangian (7). The singularity due to the term with
$\beta'$ in the denominator prevents a consideration of solutions finite at the throat a throat. This singularity
may be removed by a further transformation \cite{mexico}, and a Schr\"odinger-type equation with a regular
potential is obtained. As a result of this analysis, all previous solutions which contain throats turn out to be 
unstable, including those corresponding to black holes \cite{zhidenko}. The reason for this discrepancy is 
not clear. Perhaps, in the gauge $\delta\alpha = 2\delta\beta + \delta\gamma$, we end up with a master
equation for a pure gauge mode. Or perhaps the singularity existing in the potential of the gauge-invariant
approach has led to nonphysical modes.

\section{Black holes in $k$-essence theories}

Let us consider the Lagrangian of a $k$-essence model:
\begin{eqnarray}
          {\cal L} = \sqrt{-g}\biggr\{R - f(X) - 2V(\phi)\biggl\},
\end{eqnarray}
where $f(X)$ is a general function of the kinetic term for the scalar field,
\begin{eqnarray}
           X = \phi_{;\rho}\phi^{;\rho}.
\end{eqnarray}
For the static, spherically symmetric case, the equations of motion are:
\begin{eqnarray}
- 2\beta'' - 3\beta'^2 + 2\beta'\alpha'+ e^{2(\alpha - \beta)} &=&  e^{2\alpha}\biggr\{\frac{1}{2}f(X) + V(\phi)\biggl\},  \\
 2\gamma'\beta' + \beta'^2 - e^{2(\alpha - \beta)} &=& e^{2\alpha}\biggr\{Xf_X - \frac{1}{2}f(X) - V(\phi)\biggl\}, 
\\
 2\gamma'\beta' + \beta'^2 - e^{2(\alpha - \beta)} &=& - e^{2\alpha}\biggr\{\frac{1}{2}f(X) + V(\phi)\biggl\}, 
 \\
 \biggr[f_X e^{2\beta + \gamma - \alpha}\phi'\biggl]' &=& V_\phi.
 \end{eqnarray}

For $V = 0$, it is possible to show \cite{denis} that it only cold black holes can exist, even in the absence of 
exact solutions. To do so, the coordinate condition is chosen to be
\begin{eqnarray}
	\alpha = - \gamma,
\end{eqnarray} 
The following notations are used:
\begin{eqnarray}
         e^{2\gamma} = A, \quad e^\beta = r(u).
\end{eqnarray}
Different combinations of the equations lead to
\begin{eqnarray}
             2A\frac{r''}{r} &=& Xf_X,  \\
             A(r^2)''- A''r^2 &=& 2, \\
              2\frac{r''}{r} &=& - C^2\frac{f_X}{r^4},
\end{eqnarray}
where $C$ has the meaning of a scalar charge.
A first integral can be obtained:
\begin{eqnarray}
                \biggr(\frac{A}{r^2}\biggl)' = \frac{6m - 2u}{r^4}.
\end{eqnarray}
A horizon implies
\begin{eqnarray}
                A \rightarrow 0.
\end{eqnarray}
But, there is the relation
\begin{eqnarray}
               Af_X\phi' = \frac{C}{r^2}.
\end{eqnarray}
Regularity at the horizon requires that the components of the energy-momentum tensor should be finite, 
and since they contain $f_X$ and $\phi'$, a singularity can only be avoided if
\begin{eqnarray}
r \rightarrow \infty
\end{eqnarray}
Hence, a possible black hole must be cold.

Let us consider the particular case where
\begin{eqnarray}
f(X) = X^n.
\end{eqnarray}
The equations of motion are the following:
\begin{eqnarray}
    - 2\beta'' - 3\beta'^2 + 2\beta'\alpha'+ e^{2(\alpha - \beta)} 
		&=& \frac{\epsilon}{2}e^{2(1 - n)\alpha}\phi'^{2n} + e^{2\alpha}V,  \\
      e^{-2\alpha}[2\gamma'\beta' + \beta'^2 - e^{-2 \beta}] &=& (2n - 1 )\frac{\epsilon}{2}e^{- 2n\alpha}\phi'^{2n}
	 	- V, \\
	\beta''+ \gamma''+ \beta'(\gamma' + \beta'- \alpha') + \gamma'^2 - \alpha'\gamma'
			 &=& - \frac{\epsilon}{2}e^{2(1 - n)\alpha}\phi^{2n} - e^{2\alpha}V, \\
         (2n - 1)\phi'' + [\gamma' + 2\beta' + (1 - 2n)\alpha']\phi'  &=& 
			\frac{\epsilon}{n}\,e^{- 2(1 - n)\alpha}\,\phi'^{2(1 - n)}V_\phi.
\end{eqnarray}

A solution can be obtained for $n = 1/2$, $V = \Lambda =$ constant \cite{denis}: using the coordinate 
condition $\alpha = \gamma + 2\beta$:
\begin{eqnarray}
      ds^2 &=& \frac{\Lambda^2}{\cosh^4\theta}dt^2 - du^2
		 - \frac{\cosh^2\theta}{\Lambda^2}d\Omega^2, \\
     \phi &=& \frac{2}{3}\epsilon\Lambda( - 3\theta + 4\tanh\theta) + \phi_0, \\
      \theta &=& \frac{\Lambda}{3}u.
\end{eqnarray}
 In the quasi-global coordinate system ($\alpha = - \gamma$), the metric has the form
\begin{eqnarray}
    ds^2 &=& (1 - \Lambda\rho^2)^2dt^2 - \frac{d\rho^2}{(1 - \Lambda\rho^2)^2} 
		- \frac{d\Omega^2}{(1 - \Lambda\rho^2)}.
\end{eqnarray}
This solution seems to represent a cold black hole. But it has a very curious structure.
It is a kind of type B2 cold black hole. But, instead of having a flat asymptotic and a horizon, 
we have two horizons at $\rho = \pm 1/\sqrt{\Lambda}$, corresponding
to $u \rightarrow \pm \infty$. Beyond each horizon, we find a timelike singularity 
at $\rho \rightarrow \pm \infty$.

In Ref. \cite{denis} we also found a solution for the particular case $n = 1/3$. This solution reads
\begin{eqnarray}
ds^2 = \biggr(\frac{B_0}{k^2 x} - \frac{k^2}{2}x^3\biggl)dt^2 - \biggr(\frac{B_0}{k^2 x} - \frac{k^2}{2}x^3\biggl)^{-1}dx^2 -
\frac{1}{k^2 x}d\Omega^2,
\end{eqnarray}
where $k$ and $B_0$ are constants. This metric is completely regular at all positive $x$, including the
horizon located at $x = k^{-1}(2B_0)^{1/4}$. However, the function $f(X)$ is
singular at the horizon, in agreement with the no-go theorem proved above.

\section{Conclusions}

Scalar fields allow for obtaining exact black hole solutions, but only if the energy conditions are violated,
otherwise only naked singularities appear. This statement is valid for both minimal and non-minimal 
couplings of the scalar field with gravity. (In the only known exception \cite{kb70,bronnikov} with a 
conformal scalar field, this field is infinite at the event horizon.) A striking feature of these black holes is
that they have zero gravity and an infinite area of the event horizon. Contrary to the usual black holes, 
these cold black holes imply infinite tidal forces at the horizon. However, this is only true for massless 
fields with or without inclusion of a non-interacting electromagnetic field \cite{cbh-e}.
in full agreement with the existing no-hair theorems.

The situation changes if one includes other structures, like the string-inspired dilatonic interaction between 
scalar and electromagnetic fields \cite{br-sh77,clement1,clement2}: in this case, a more conventional 
black-hole horizon can appear, with finite area and finite temperature, in the presence of a usual scalar field. However, if the scalar field is phantom, there emerge cold
black hole solutions \cite{clement1,clement2}. There are also black hole solutions, albeit numerical, with "true" scalar hair (i.e. an independent scalar charge which is conserved) in Einstein's gravity minimally coupled to two real (or one complex) massive scalar fields, and thus obeying all energy conditions \cite{radu}.

The stability issue looks somewhat controversial (see above), but most probably the gauge (44), in which 
scalar field perturbations decouple from the metric ones, excludes the physical unstable mode found in
\cite{mexico,zhidenko}.

$k$-essence models give origin to similar structures, but with some other exotic configurations, 
to be still better explored. In particular, we find, at least for some particular cases, multiple horizons 
beyond which infalling observers hit singularities. In such cases the usual notion of a distant observer is 
inapplicable since there is no spatial infinity. The features of such solutions, including their stability, 
deserve more complete studies.

We here did not touch upon a broader area of possible black holes with scalar hair in cases where scalar
fields possess nontrivial self-interaction potentials. In such cases, apart from general results like different 
no-hair and other no-go theorems (see, e.g., \cite{adler,bek,kb01,radu}), there are a large number of exact solutions of
interest, including singular and regular black holes with normal and phantom scalar fields with and without
electromagnetic fields, see, e.g., \cite{BR, we-prl, bh-bu, bu-e} and references therein.

\section*{Acknowledgments}

We thank CNPq (Brazil) and FAPES (Brazil) for partial financial support.
The work of KB was partly performed within the framework of the Center 
 FRPP supported by MEPhI Academic Excellence Project (contract number 02.a03.21.0005, 27.08.2013).

\end{document}